\documentclass[12pt,a4paper]{article}
\usepackage[utf8]{inputenc}
\usepackage[english]{babel}
\usepackage{amsmath}
\usepackage{amsfonts}
\usepackage{amssymb}
\usepackage{makeidx}
\usepackage{graphicx}
\usepackage{morefloats}
\usepackage[left=2cm,right=2cm,top=2cm,bottom=2cm]{geometry}
\author{Arindam Ghosh$^{1}$ and Sandip K. Chakrabarti$^{1,2}$}
\title{Periodic X-ray Modulation and its Possible Relation with Orbital Elements in Compact Binaries}
\begin{document}
\maketitle
\begin{center}
{$^1$S. N. Bose National Centre for Basic Sciences, Salt Lake, Kolkata 700106, India.\\ 
$^2$Indian Centre for Space Physics, Chalantika 43, Garia Station Rd., Kolkata 700084, India.\\
{\it arindam.ghosh@bose.res.in; sandip@csp.res.in}\\}
\end{center}
\abstract{Stellar companion of a black hole orbiting in an eccentric orbit will experience modulating tidal force 
with a periodicity same as that of the orbital period. This, in turn, is expected to modulate 
accretion rates, and thus the seed photon flux which are inverse Comptonized to produce harder X-rays
are expected to be modulated. However, there could be several other effects, such as eclipses, reflection from winds,
superhump phenomena etc., which could also modulate the radiation at about the same frequency.
By analyzing complete all sky monitor (ASM) data ($1.5-12$ keV) of RXTE 
and all sky survey data ($15-50$ keV) of Swift/BAT satellites,
we find presence of this periodicity in several objects. If we assume that this modulation is
solely due to tidal effects, the RMS-value of the peak in power density spectrum allows us to estimate eccentricities 
of these orbits. Surprisingly, these agree with eccentricities measured by independent means for 
some of these objects. For the rest, our estimate of eccentricities may be taken as the upper limit.}\\ 


\section{Introduction}

X-ray flux from stellar mass black holes is generally known to be time dependent. Apart from strong temporal and
spectral variations which are manifested by spectral state transitions and possibility of quasi-periodic oscillations
(e.g., Chakrabarti \& Titarchuk, 1995), there could be subtle modulations which contain information about
orbital periodicies. Using two year of RXTE/ASM data for Cyg X-1, Wen et al. (1999) detected the
$5.6~d$ orbital period in Lomb-Scargle periodograms of both light curves and hardness ratios when 
Cyg X-1 was in the hard state though this feature was `absent' in the soft state. 
However, Boroson and Vrtilek (2010) found orbital variability in Cyg X-1 during the soft states
using light curves provided by the long-time RXTE/ASM data with a similar technique leading to their
conclusion that the orbital variability in all soft states could be detected if the span of data 
was sufficiently longer. Wen et al. (1999) argued that absorption of X-rays by a stellar wind from the
companion star could reproduce the observed X-ray orbital modulations in the
hard state. They suggested that a reduction of the wind density be required in 
order to explain the low orbital modulation in the soft-state data.
Later Wen et al. (2005) analyzed long-time (8.5 years) RXTE/ASM data using 
the same periodogram technique and showed periodic modulation in a large number 
of X-ray  sources. Their systematic analysis revealed that orbital 
modulation was more readily detected in HMXBs than in LMXBs. 
The fraction of eclipses from their observations in LMXBs 
was $<3\%$, which is much less compared to that of the HMXB systems. 

Of course, this is not thought to be the only cause of modulation seen in a long term X-ray data. 
Patterson et al. (2005, and references therein) reports that cataclysmic binaries exhibit 
superhumps with periodicity a few percent higher than the orbital period provided 
the mass ratio $q=M_2/M_1<0.3$ (where, $M_2$ and $M_1$ are the companion mass and the 
compact mass respectively). Boyd, Smale and Dolan (2001) find X-ray modulation in LMC X-3 at the known orbital period
of $1.7$ days. Brocksopp et al. (1999a) find a similar modulation in Cyg X-1. Smith et al. (2002a), 
using five years of RXTE data of galactic black hole candidates 1E 1740.7-2942 and GRS 1758-258, 
show that these have periodic modulations of $12.73 \pm 0.05$ days and $18.45 \pm 0.10$ days respectively and interpreted these
as the orbital modulations. Later, Obst et al. (2013) found this number to be drifting for GRS 1758-258. 
Kudryavtsev et al. (2004) reported obtaining several such periodicities in the MIR satellite data and they identified
some of these with periodicites of known objects, such as, H 1705-25, GRO J1655-40, and 4U 1543-47.
They did not identify the reasons of such a modulation but concluded that these periodicities may not be connected to eclipse.

There could be another cause of periodic modulation, particularly if the orbit has non-zero eccentricity,
howsoever small. This is likely to be true, since an exact circular orbit could be impossible to achieve as that requires 
many fine tuning. In such an orbit, tidal force $F_t$ exerted by the primary on the companion would be inversely 
proportional to the cube of the distance between the two components. If semi-major axis and eccentricity 
are $a$ and $e$ respectively, then 
$$
F_{t,a} \propto \frac{1}{[a(1+e)]^3}, 
\eqno{(1a)}
$$
and
$$
F_{t,p} \propto \frac{1}{[a(1-e)]^3},
\eqno{(1b)}
$$
would be the net tidal force at apoastron (marked by subscript `t,$a$') and periastron (marked by subscript `t,p') respectively 
(Bradt, 2008). In our context, where we consider the passage of a companion 
around a black hole, we use the terms aponigrumcavum  
and perinigrumcavum (ANC and PNC in short) to represent apoastron and periastron respectively 
(`nigrum cavum' being `black hole' in Latin). Supply of matter from the companion 
could be assumed to be proportional to this tidal effect and that will have a periodic variation 
in the flow rate. Ratio of any of the component rates at ANC and PNC would be,
$$
\frac{{\dot M}_a}{{\dot M}_p} = \frac{{\cal F}_a}{{\cal F}_p} = \frac{(1-e)^3}{(1+e)^3} ,
\eqno{(2)}
$$
where, ${\cal F}$ represents radiation fluxs at respective locations. The first equal sign is valid for a Keplerian 
disk if eccentricity is small enough so that the Roche lobe remains filled at both periastron and apoastron.
In this case, the modulation of the Keplerian rate 
will propagate in viscous time scales and would enhance soft X-rays which act as seed photons.
This will, in turn, increase the flux of Comptonized X-rays. We should therefore expect a modulation of X-rays 
at orbital period in all the compact binaries whose eccentricity is non-zero.

It is well known that spectral properties of black holes cannot be explained by a 
Keplerian disk alone. Along with a standard Keplerian disk, 
a hot electron component is required (Zdziarski 1988; Haardt et al. 1994; Zhang et al. 2009).
In Chakrabarti \& Titarchuk (1995) solution, this hot component is created by a low angular momentum, 
radiatively inefficient flow, which slows down close to the black hole due to centrifugal forces 
and puffs up by resulting heat. This so-called CENtrifugal pressure supported BOundary Layer, 
or CENBOL, behaves as the Compton cloud and reprocesses low energy (soft) photons 
into high energy (hard) photons. Detailed analysis of actual satellite data revealed 
that this so-called two component advective flow solution can explain even subtle aspects 
of spectral and timing properties of black hole candidates including outflows and quasi-periodic oscillations
(Chakrabarti \& Manickam 2000; Rao et al. 2000; Smith et al. 2001, 2002a,b, 2007; Wu et al. 2002; Debnath et al. 2010; 
Soria et al. 2011; Nandi et al. 2012; Cambier \& Smith 2013). General success of such model indicates
that the Compton cloud is also produced by the Companion and understanding of hard X-rays do 
not require any external source of electrons. However, a sub-Keplerian flow, arising out of winds 
is also expected to be modulated and this flow would arrive in almost free-fall time scale if viscosity is low enough so that 
it does not become a Keplerian disk. However, as Chakrabarti \& Titarchuk (1995) found, 
that would only modulate the optical depth of the `Compton cloud' and will change spectral 
slopes without changing the photon flux significantly.
Smith et al. (2001, 2002b) pointed out that the RXTE/ASM data points to 
the existence of two components in the accretion flows. They find that there is a distinct 
time-lag between photon index and flux in low mass X-ray binary systems which 
accrete primarily through the Roche Lobe overflow, while high mass X-ray binaries, 
which primarily accrete winds of the companion, does not show such a lag. 

In the present paper, we explore this aspect and show that indeed the majority 
of X-ray binaries having a black hole exhibit such a modulation 
with a periodicity exactly or nearly the same as the orbital time period (if known). 
The total modulation clearly depends on several factors, such as, mass of the
compact object, size and mass of the companion, mass loss due to winds, probable eclipsing 
or reflection by the atmosphere of or winds from the companion, superhumping,
etc. all of which may not be known with certainty as there are very little theoretical 
progress in each of these directions. However, the modulation due to tidal force variation
can be quantified and the RMS value of the peak in power density spectrum could be used to 
limit the upper limit of the eccentricity. We therefore concentrate on the flux ratio at ANC 
and PNC which depends only on eccentricity (Eq. 2). We use publicly available 
RXTE/ASM and Swift/BAT data from all sky survey instruments. 
We show using Fourier analysis of light curves, and Lomb-Scargle type
periodograms that the modulation at quasi-orbital period (QOP) is real. 
From the rms value of the modulation, we estimate the orbital eccentricity as well.  
In future, we will consider spectral properties and how they are modified by these tidal effects.
In the next Section, we present our data analysis technique. In Section 3, we discuss known 
properties of several stellar mass black hole candidates which are relevant to our work.
In Section 4, we present results of our analysis. Finally, in Section 5, we draw our conclusions.

\section{Data Analysis}

We use {\sc ascii} versions of public/archival RXTE/ASM dwell-by-dwell light curve data 
(MJD 50455 onwards) and Swift/BAT orbital light curve data (MJD 53415 onwards) 
for the X-ray binary sources Cyg X-1, Cyg X-3, XTE J1650-500, H 1705-25, 1E 1740.7-2942 and GRS 1758-258. 

The RXTE/ASM has a collecting area of $90$ cm$^2$ and is operating over $1.5-12~keV$ range.
The Swift/BAT has an all-sky hard X-ray surveyor. It is operating over $15-150$ keV
with a detecting area of $5200$ cm$^2$. We have used the data inside $15-50$ keV energy range. 
In order to observe long-time behaviour of the aforesaid sources, RXTE/ASM 
data for about $13$ years and Swift/BAT data for over $8$ years without any truncation, are used. 
The standard {\sc ftools} package of {\sc heasoft} (Version 6.13) is
used for data reduction.
Neither of the instruments are meant to obtain data continuously from any particular source,
so there are `gaps' in the data, especially, due to annual solar constraints. In order to 
overcome these shortcomings, particularly unevenness of data interval, we use a {\sc fortran} code 
for interpolating data at equal time intervals, required for carrying out Fourier
analysis. Using the Perl script {\sc ascii2flc}, the {\sc fits} files for the light curves 
suitable for {\sc xronos} are created. These are used to produce power density spectra (PDS) using {\sc powspec} 
task of {\sc xronos} (Version 5.22) package. In order to reduce noisy background and low frequency noise,
we take running average of evenly spaced data and subtract it from the initial data. 
Fluctuations of data around the running mean are then considered for producing PDS. 
Normalization factor of $-1$ is taken in {\sc powspec} task. We extracted white-noise-subtracted spectra for time
bins of $0.01$ d, $0.025$ d and $0.05$ d - or $14.4$ min, $36$ min and $72$ 
min respectively. Since observed peaks are expected to be Lorentzian type, PDS are fitted with 
Lorentzian profiles in order to determine centroids frequencies. 
The centroid frequency is considered to be due to the quasi-orbital period (QOP).
We also estimated the Q-factor (centroid frequency/width of the
Lorentzian), errors, and the significance of the observed QOPs using the {\sc fit err} and {\sc uncer} commands. All the QOP frequencies are estimated at 90\% confidence level ($1.64\sigma$). The standard {\sc qdp/plt} command, {\sc statistics}, 
is used to obtain the area under a fitted peak when zoomed above the red noise level 
(integrated rms power for a QOP), $\chi^2$ values and degrees of freedom (DOF). 
Standard conversion tables of  $\chi^2$/DOF distributions, p-values and Gaussian probabilities give 
the statistical significance of the area in terms of the standard deviation ($\sigma$) of 
Gaussian distribution. It may be noted here that the eccentricity of a given orbit 
depends solely on the area under the fitted peak of a given PDS.
Non-constancy of viscosity is expected to reduce the sharpness of the peaks and thus
we would observe a relatively flatter peak at QOP. For convenience, we choose 
units of frequencies as $d^{-1}$ instead of $Hz$. We also plot light curves with actual data in modified Julian day (MJD).  

In addition to using our own code for interpolating data before getting PDS, we also make periodograms 
of Lomb-Scargle type from the aforesaid fluctuations of all the sources. This is for an additional check to see if the QOPs obtained from PDS are also reflected in 
the periodograms. We have examined p-values and phase diagrams in all cases, though statistical significance 
of the peaks is often questioned. In this period step method, we choose fixed step sizes of $0.01$ d, 
$0.001$ d, $0.005$ d, $0.0005$ d, $0.0025$ d and $0.00025$ d depending on the nature of the data and 
the periodicity that we wish to focus on. But it should be stated here that all our results are based on the outcome of the PDS only.


\section{X-ray Binaries under Consideration}

\subsection{Cyg X-1}

Cyg X-1 system is believed to have a stellar mass black hole and a blue supergiant star forming a binary. 
Orbital period is $5.599829\pm0.000016 d$ (Brocksopp et al. 1999b). The system does not eclipse. 
The orbital eccentricity of $0.018\pm0.003$ is very small giving rise to a nearly circular orbit
(Orosz et al. 2011) or it is appreciably high with another estimation of $0.06\pm0.01$ (Bolton 1975).  
There is some uncertainty about the mass of the compact object and its companion. 
Stellar evolutionary models suggest a mass of $21\pm8~M_{\odot}$ turning around an O$9.7$Iab companion of 
$40\pm10~M_{\odot}$ (Zi\'{o}\l kowski 2005) while other techniques resulted in  $M_{1}=10M_{\odot}$. 
Quasi-Periodic Oscillations (QPOs) in X-ray emission has yielded a more precise value of mass
$14.8\pm1~M_{\odot}$ and a companion mass of $M^{opt}_{2} = 19.2 \pm 1.9~M_{\odot}$ (Orosz et al. 2011). 
INTEGRAL/ESA's view of Cyg X-1 reveals that the companion star (HDE 226868) is a supergiant with an estimated 
surface temperature of $31000~K$ and a mass of approximately $20-40~M_{\odot}$. 
Based on a stellar evolutionary model, at the estimated distance of $2~kpc$, the companion may have a 
radius of about $15-17~R_{\odot}$ (Orosz  et al. 2011) 
and has a luminosity of approximately $3-4\times10^{5}~L_{\odot}$ (Zi\'{o}\l kowski 2005). 
The compact object is estimated to be orbiting its companion at a distance of about $40~R_{\odot}$ (Miller et al. 2005) 
and the system has an inclination $i=27.1^{\circ}\pm0.8^{\circ}$ (Orosz et al. 2011).

\subsection{Cyg X-3}
Cyg X-3 is an accreting X-ray binary with a relativistic jet, observed from radio to high-energy gamma-rays. 
It consists of a black hole of mass $1.3-4.5~M_{\odot}$ (Zdziarski et al. 2013) which is wind-fed by a Wolf-Rayet star. 
Its distance from us is about $\sim7~kpc$. It is a high-mass X-ray binary as the companion (V1521 Cyg) is a high-mass star, having a 
mass of $7.5-14.2~M_{\odot}$. The orbital period is about $4.8~h$ (Parsignault et al. 1972; Davidsen \& Ostriker 1974)  with an
inclination of $34^{\circ}-54^{\circ}$ (Zdziarski et al. 2013). Binary separation is $d\sim3\times10^{11}~cm$. 
Only partial eclipses are observed in its otherwise obscured character at low energy 
(ROSAT home page). The companion has a strong wind ($\dot{M_{w}}\sim10^{-5}M_{\odot}~yr^{-1}$, $v_{w}\sim1000~km s^{-1}$)
(Dubus et al. 2010a). Scattering in the wind washes out rapid X-ray variability timescales 
and also modulates X-ray emission. Based on a mass-independent model of variable luminosity X-ray 
source, and an elliptic orbit, the system was reported to have an eccentricity of  $\sim0.14$ and 
a smaller orbital inclination ($i=24^{\circ}$) (Ghosh et al. 1981). However, this limit would constrain 
$M_{1}<3.6M_{\odot}$ and the companion to $M_{2}< 7.3M_{\odot}$ (Stark \& Saia 2003). 
Fermi observations show that high energy gamma-ray flux is modulated with the orbital period. 
Gamma-ray modulation is almost in anti-phase with X-ray modulation (Dubus et al. 2010b).

\subsection{XTE J1650-500}

XTE J1650-500 is a stellar mass black hole candidate probably having a mass of $3.8\pm0.5~M_{\odot}$
(Shaposhnikov \& Titarchuk 2009). A safer mass range is $2.7-7.3~M_{\odot}$ (Orosz et al. 2004).
The binary period is $7.63~h$ and an estimation of the inclination is $50^{\circ}\pm3^{\circ}$ (Orosz et al. 2004).

\subsection{H 1705-25}

It is a binary system with a small and cool star having about $0.3~M_{\odot}$ (Orosz \& Baily 1997) 
as a companion to the black hole. The orbital period is
about $12.5~h$ (Johannsen et al. 2009). Detection of an orbital and ellipsoidal modulation 
at a period of $16.8~h$ with an orbital inclination in the range $48^{\circ}<i<51^{\circ}$ and 
distance $2-8.4~kpc$ were also reported earlier (Martin et al. 1995). 
From the characteristics of the companion star and its orbit, estimated mass of the black 
hole is $6\pm2~M_{\odot}$ (Johannsen et al. 2009).

\subsection{GRS 1758-258 and 1E 1740.7-2942}

GRS 1758-258 is often referred to as a twin source along with 1E 1740.7-2942. Both are persistent sources 
above $\sim50$ keV and are located in the vicinity of the Galactic centre.
They emit X-rays and display relativistic jets in the radio
band. The X-ray luminosity, hard spectra, persistent activity and the shape of the power spectra in  
1E 1740.7-2942 and GRS 1758-258 have made these two objects comparable to Cyg X-1 (Main et al. 1999). 
Remoteness and high column density towards GRS 1758-258 does not allow observations below ($25~keV$). 
Search for a counterpart in optical and infrared has not turned up any distinct result but 
only two or more candidates were found within $1''$. Weak jet structure reveals that GRS 1758-258 and 
1E 1740.7-2942 are microquasars (Keck et al. 2001).

The mass of the companion of the black hole in GRS 1758-258 is $M_2\leq4M_{\odot}$ and 
in 1E 1740.7-2942 is  $M_2\leq9M_{\odot}$. The presence of a well-collimated radio 
jet in 1E 1740.7-2942 is an indicative of a stable accretion disk. Companion of either of these two sources 
is not able to feed the black hole via its stellar wind and they are powered by binary accretion via 
Roche lobe overflow. They are LMXBs with companion mass $\sim1~M_{\odot}$. Apparent association of the source 
with a high-density molecular cloud hints at a possible accretion directly from the ISM. 
The binary period in both cases is likely to be shorter than $20~h$ (Chen et al. 1994). 
Assuming that the optical component in GRS 1758-258 is a main-sequence star and $M_1\geq3~M_{\odot}$ 
\& $M_2<4~M_{\odot}$, the period is $\leq5.8~h$ and $M_2\sim0.65~M_{\odot}$ (Kuznetsov et al. 1999).


\section{Results of our Analysis}

Figure 1 shows the power density spectra (PDS) of the reduced data obtained from the fluctuations around running average from the light curves of the six sources as obtained from the ASM data ($1.5-12$ keV) for 5000 days of RXTE as discussed in Sec. 2. we show light curves of the same objects,  but drawn using . Data duration is about 3000 days. The PDS extracted from the all sky survey data for 3000 days of Swift/BAT 
($15-50$ keV) after subtracting the dynamic mean 
are shown in Figure 2. The time scales corresponding to QOPs are written from the Lorentzian fitting. 
The reason for the bumps at  $\sim10^{-6}Hz$ in the PDS of Cyg X-1, 1E 1740.7-2942 \& 
GRS 1758-258, and at $\sim10^{-5}Hz$ in the PDS of Cyg X-3, XTE J1650-500 \& H 1705-25 is inherent 
to the procedure or removal of low-frequency noise as discussed in Sec. 2.  

Although all the Lorentzians are fitted for $90\%$ confidence level ($1.64\sigma$), the peaks of Fig. 2 (with Swift data) are appreciably more distinct and unique over 
those of Fig. 1 (with RXTE data). Interestingly, the periods obtained from both RXTE and Swift data are in the same ball-park. Distinct peaks at $\sim5.6~d$ and $\sim4.8~h$ are observed for Cyg X-1 and Cyg X-3, respectively, among others. The prominent peak of Cyg X-3 is not due to its
eclipsing nature  since eclipsing is strong only at low energies, and not in Swift/BAT energy range. 
The periodicities of two enigmatic black holes GRS 1758-258 and 1E 1740.7-2942 are not yet known correctly, 
their apparent periodicities are around $\sim 3-4~d$ as seen in our analysis. 
In order to have an additional check on the periodicities of the obtained PDS, 
we draw Lomb-Scargle type periodograms for all these objects. In Fig. 3, we show 
periodograms drawn for fluctuations around the mean on the left column  for RXTE/ASM data and
on the right column  for Swift/BAT data. Though several distinct peaks are seen in 
many of these data, we see evidence of periodicities having values similar to what 
we obtained through PDS. The periods marked in each box are obtained from the strongest peaks. The strongest peak of GRS 1758-258 in Swift data is double, possibly because of the superhumping. Similarly, the RXTE/ASM also shows a double peak at around $T=12h30m$. Although we have shown periodograms, we wish to reiterate that our results only rely on the PDS as far as the extracted 
time periods are concerned. The eccentricities are obtained only with PDS from the Swift, 
as the area of the detector is large giving rise to more reliable results in high energies. This makes the 
peaks to be more distinct over those obtained from RXTE/ASM data.

Table 1 summarizes results of our analysis. Known system parameters are given with 
references. In Column 1, names of the systems are written. In Cols. 2 and 3, we present known
masses of black holes and companions. In Cols. 4 and 5, we present inclination 
angles and periods as reported in the literature. In Cols. 6 and 7, we present centroid 
periods $T$ as obtained from power density spectra (PDS) of the complete data 
set from Swift (Col. 6) and RXTE (Col. 7) after fitting peaks with 
Lorenztian as described earlier. In Cols. 8 and 9 we present periods obtained from Lomb-Scargle periodogram 
from these two satellites. Finally, in Col. 10, on the top, we present estimated rms power (in \%), and in the middle,
eccentricity ($e$) of the orbits. Also, a measure of statistical significance (in units of standard 
deviation $\sigma$ of Gaussian distribution) is given at the bottom (within parentheses).

The procedure to obtain eccentricity is the following. 
We obtained {\it rms} power of the Lorentzian which fits the periodicity. For high mass X-ray 
binaries (HMXBs), such as Cyg X-1 and Cyg X-3, the rms powers obtained from the fitted Lorentzian 
are found to be $6.72\%$ and $5.30\%$ respectively. These are consistent with typical 
estimates of {\it rms} power (Choudhury et al. 2004).  From Eq.(2), if we assume that the 
ratio is proportional to the ratio of the number of soft photons emitted at ANC and PNC
then one can derive the fractional change in photon counts $\frac{\Delta N}{N}$ 
of {\it rms} fraction as a function of the eccentricity $e$ as,
$$
\frac{\Delta N}{N} = \frac{3e+e^3}{1+3e^2}. 
\eqno{(3)}
$$
Using Eq.(3), we obtain the eccentricities as $0.0224$ for Cyg X-1 and $0.0177$ for Cyg X-3 respectively. Our estimate of 
eccentricity for Cyg X-1 ($0.022$) lies between earlier two estimates of $0.018-0.025(\pm0.003)$ 
(Orosz et al. 2011) and $0.06\pm0.01$ (Bolton 1975) but nearly equals to the former. 
Although $e=0.018$ of Cyg X-3 is much lower than the old reports (Sec. 3.2), it is consistent with 
that of Cyg X-1. Both the HMXBs have low angles of inclination but their modulation is high. 
Cyg X-1 does not eclipse at all. Although partial eclipses are reported to be observed 
in Cyg X-3, but the system is heavily obscured at low energies. 
The conclusion we draw from our results is that the modulations causing the peak in the PDS are  due to tidal effects. 
Upper limit of inclination angles of H 1705-25 and XTE J1650-500 is $\sim50^{\circ}$ and so far there is no available report 
about eclipsing or wind reflection effects. 
This leads us to believe that other $e$-values which we presented in Table 1 may represent 
actual eccentricities of the respective orbits. In presence of other effects, these numbers could be treated 
as the upper limits.

\section{Discussions and Conclusions}

In accreting X-ray binaries, X-ray flux modulations in orbital scales 
are common phenomena. So far, orbital modulations are attributed 
to eclipsing effects, effects from wind reflections and sumperhumping. From an well established
theoretical argument of dependence of tidal effects on eccentricity
we suggest that the mass transfer and the orbital modulation would also be affected for those
binaries which have non-zero eccentricity. Smaller distance during a perinigrumcavum passage would 
enhance mass accretion rates and larger distance during an aponigrumcavum passage would reduce it. 
In case the eccentricity is not high, the binary remains accreting for the entire orbit. 
Thus photons emitted from the Keplerian disk is expected to be modulated in a single 
orbital period. However, viscous time scale would introduce a time lag to 
these modulations and the time lag would be constant for constant viscosity.
Viscosity may vary with time causing a spread in the modulation. Thus a Fourier analysis of the light curve
is expected to yield a quasi-orbital periodicity (QOP). There could be many factors, such as, 
non-locking of spin and orbital angular momentum of the companion, large variation of viscosity, 
possible magnetic effects (which may enhance disk accretion rates), eclipsing of 
Compton clouds, which can wash out such a modulation. However, we still find in several objects that modulation period is 
almost same as the estimated time periods obtained from other methods, at least in those cases
where they are known. For HMXBs, the {\it rms} values at the peak of power density spectra gives directly 
$e$, assuming that the modulation is solely due to tidal effects. Surprisingly, 
for Cyg X-1, for which the eccentricity was known from previous studies, our result agrees 
very well. For Cyg X-3, eclipsing is known to take place, though it may not affect the
high energy observations by Swift. For all the cases, the estimates are presented in Table 1. 

Properties of emissions from accretion disks around black holes are largely sensitive to outer boundary 
conditions of the disks. These in turn, depend on properties of the companion. We therefore need 
to understand long term properties of mass-loss of the companion if we wish to predict X-ray 
properties beyond a fraction of the orbital period in which many things can happen to the supply rate of matter.
A random variation of viscosity, for instance, can change the  
accretion rates in Keplerian and sub-Keplerian components at various time scales. This, together with the tidal effects, may 
wash out any periodicity in the light curves. It is thus surprising that we do see this quasi-orbital periods
in PDSs or periodograms even when we analyze several years of data together. What is most astonishing is that X-rays which are
supposed to be emitted from the very inner region of a disk, carries this information on tidal force modulation on the companion. 

It is often believed that X-rays may be modulated by eclipsing effects of winds or reflections from the winds. However, in 
computing spectral fits, it is universal policy not to include effects of such eclipse or subtract effects of the reflections
from the winds of the companions at certain phases. This testifies that these effects cannot be important in understanding 
spectral information. This strengthens our belief that the effects due to tidal force could be important and our procedure 
could be extended to all the systems having non-zero eccentricity. As discussed already, we anticipate a similar modulation in spectral index, since 
the sub-Keplerian winds would also be modulated at a different phase in an orbital period. 
This study will be reported elsewhere.

\section*{Acknowledgements}
The authors thank NASA Archive for RXTE/ASM and Swift/BAT data and Nasa Exoplanet Archive for producing periodograms.

\begin{table}

\caption{\bf Estimated Orbital Parameters of Six X-Ray Binaries}
\footnotesize\rm
{\tiny
\begin{tabular}{c|ccccccccc}
\hline\noalign{\smallskip}
\hline
&&&&&&\\
X-Ray Binary & $M_{1}$ &  $M_{2}$ & Inclination &  Period & \multicolumn{2}{c}{T (PDS)}&\multicolumn{2}{c}{T (LS Periodogram)}& rms (\%)\\
(BH/BHC)&$(M_{\odot})$&$(M_{\odot})$&$(i^{\circ})$& (T) & ({\it Swift})& ({\it RXTE})& ({\it Swift})& ({\it RXTE}) & {\bf $e$}\\
&&&&&&&&& (Stat. Sig.)\\
\hline
&&&&&&&&&6.72\\
Cyg X-1 & $14.8\pm1^a$ & $19.2\pm1.9^a$ & $27.1\pm0.8^a$ &  $5.6~d^b$ &$5.53^{+0.70}_{-0.56}~d$&
$5.60^{+0.33}_{-0.30}~d$ & $5.60~d$ & $5.60~d$  & 0.0224\\
&&&&&&&&&$~(3.3\sigma)$\\
\hline
&&&&&&&&&5.30\\
Cyg X-3 & $2.4^{+2.1c}_{-1.1}$ & $10.3^{+3.9c}_{-2.8}$ & $34-54^c$ &  $4.8~h^{c,d}$ & $4.87^{+0.23}_{-0.21}~h$ & $4.86^{+0.16}_{-0.15}~h$& $4.80~h$& $4.80~h$ & 0.0177\\
&&&&&&&&&$~(1.6\sigma)$ \\
\hline
&&&&&&&&&3.15\\
H 1705-25 & $6\pm2^e$ & $0.3^f$ &$48-51^g$ & $12.5~h^e$ & $12.47^{+1.09}_{-0.92}~h$ & $12.70^{+1.48}_{-1.20}~h$ & $12.45~h$ & $12.30~h$ & 0.0105\\ 
&&&&&&&&&$~(1.6\sigma)$\\
\hline
&&&&&&&&&3.21\\
XTE J1650-500 & $2.7-7.3^{h,i}$ & - & $50\pm3^h$ &  $7.63~h^h$ & $7.82^{+0.83}_{-0.68}~h$ & $7.49^{+0.79}_{-0.65}~h$&$7.54~h$ & $7.84~h$  & 
0.0107\\
&&&&&&&&&$~(2.6\sigma)$ \\
\hline
&&&&&&&&&2.49\\
GRS 1758-258 & $\geq3^j$ & $0.65^j$ & - &  $\sim6~h^j$ & $3.61^{+0.21}_{-0.19}~d$& $3.38^{+0.27}_{-0.23}~d$ & $3.60~d$ & $3.42~d$ 
& 0.0083\\
&&&&&&&&&$~(2.0\sigma)$ \\
\hline
&&&&&&&&&4.10\\
1E 1740.7-2942 & - & $1-9^k$ & - &  $<20~h^k$ & $3.74^{+0.72}_{-0.52}~d$ & $ 3.21^{+0.48}_{-0.37}~d$ & $3.69~d$ &$3.66~d$ &
0.0137\\
&&&&&&&&&$~(1.3\sigma)$\\
\hline\noalign{\smallskip}
\hline
\end{tabular}}

\noindent$^a$Orosz et al. (2011), $^b$Brocksopp et al. (1999b), $^c$Zdziarski et al. (2013), $^d$Davidsen \& Ostriker (1974), $^e$Johannsen et al. (2009), $^f$Orosz \& Baily (1997), $^g$Martin et al. (1995), $^h$Orosz et al. (2004), $^i$Shaposhnikov \& Titarchuk (2009), $^j$Kuznetsov et al. (1999), $^k$Chen et al. (1994).
\end{table}

%

\begin{figure}
\begin{center}
\includegraphics[height=14 cm, angle=0]{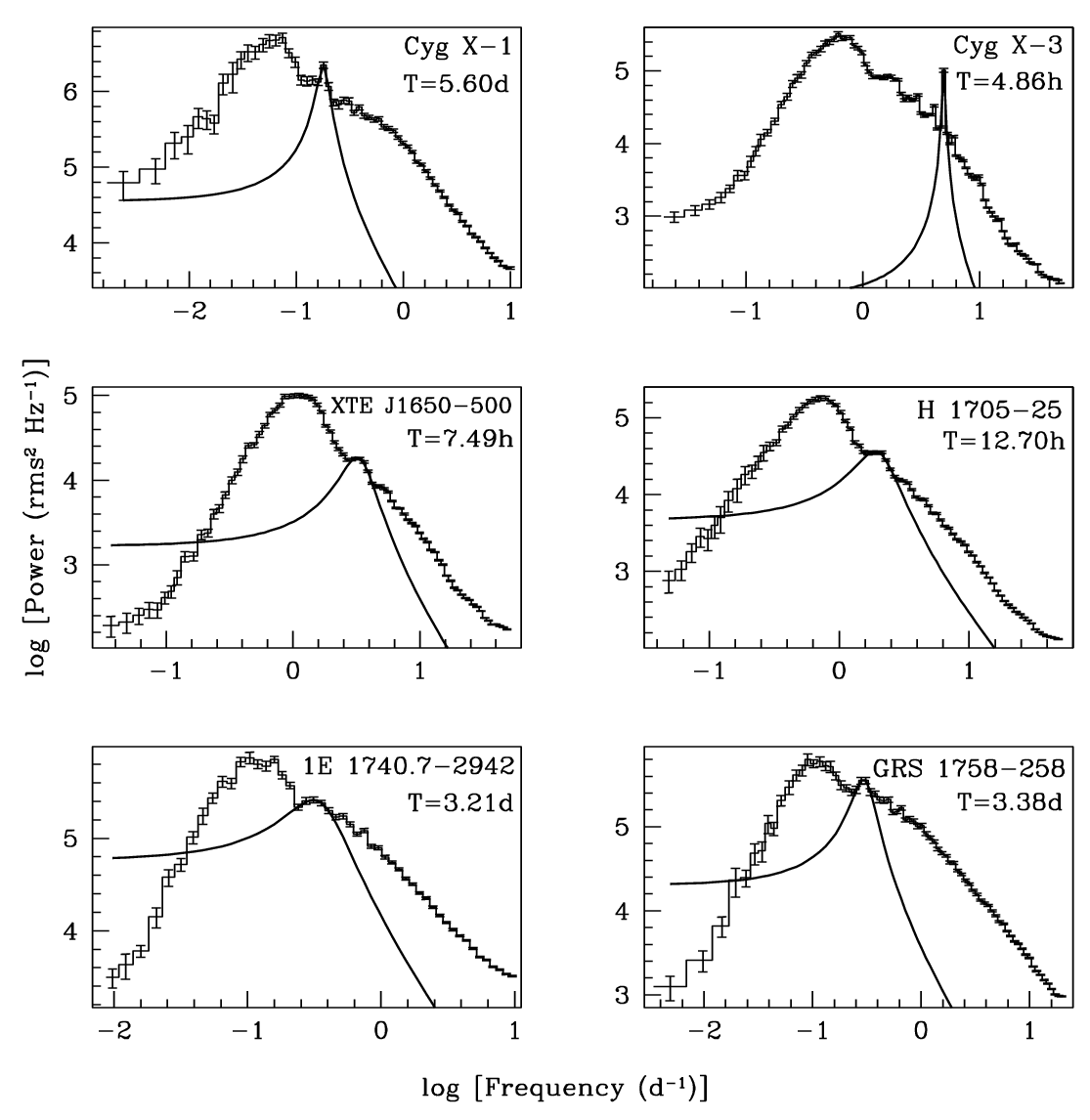}
\caption{Power Density Spectra with modified RXTE/ASM data. Peaks due to QOPs are fitted with Lorentzians. Centroid periods are marked in each box.}
\end{center}
\end{figure}

\begin{figure}
\begin{center}
\includegraphics[height=14 cm, angle=0]{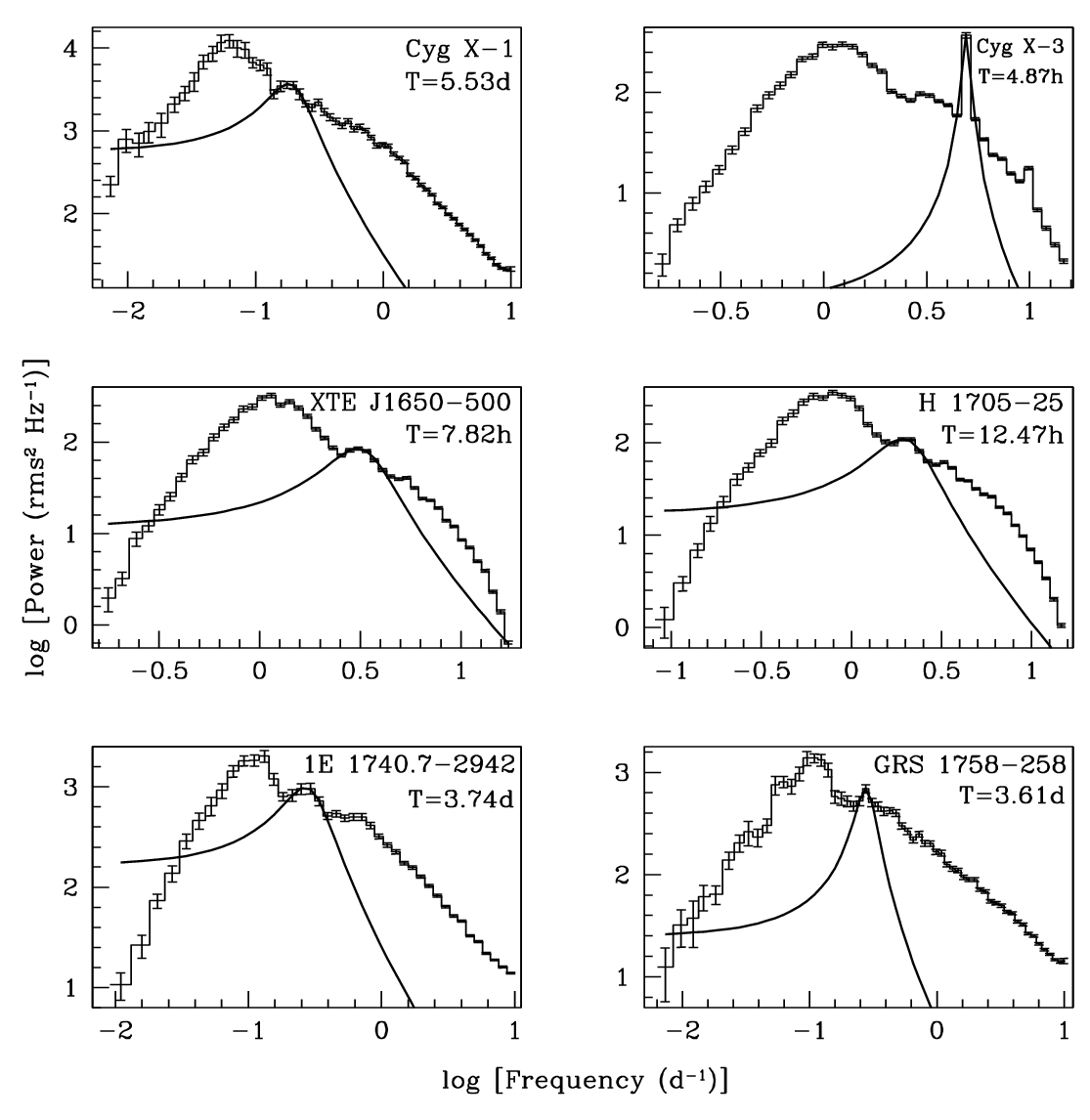}
\caption{Power Density Spectra with modified Swift/BAT data. Fitted QOPs are marked in each box.}
\end{center}
\end{figure}


\begin{figure}
\begin{center}
\includegraphics[height=14 cm]{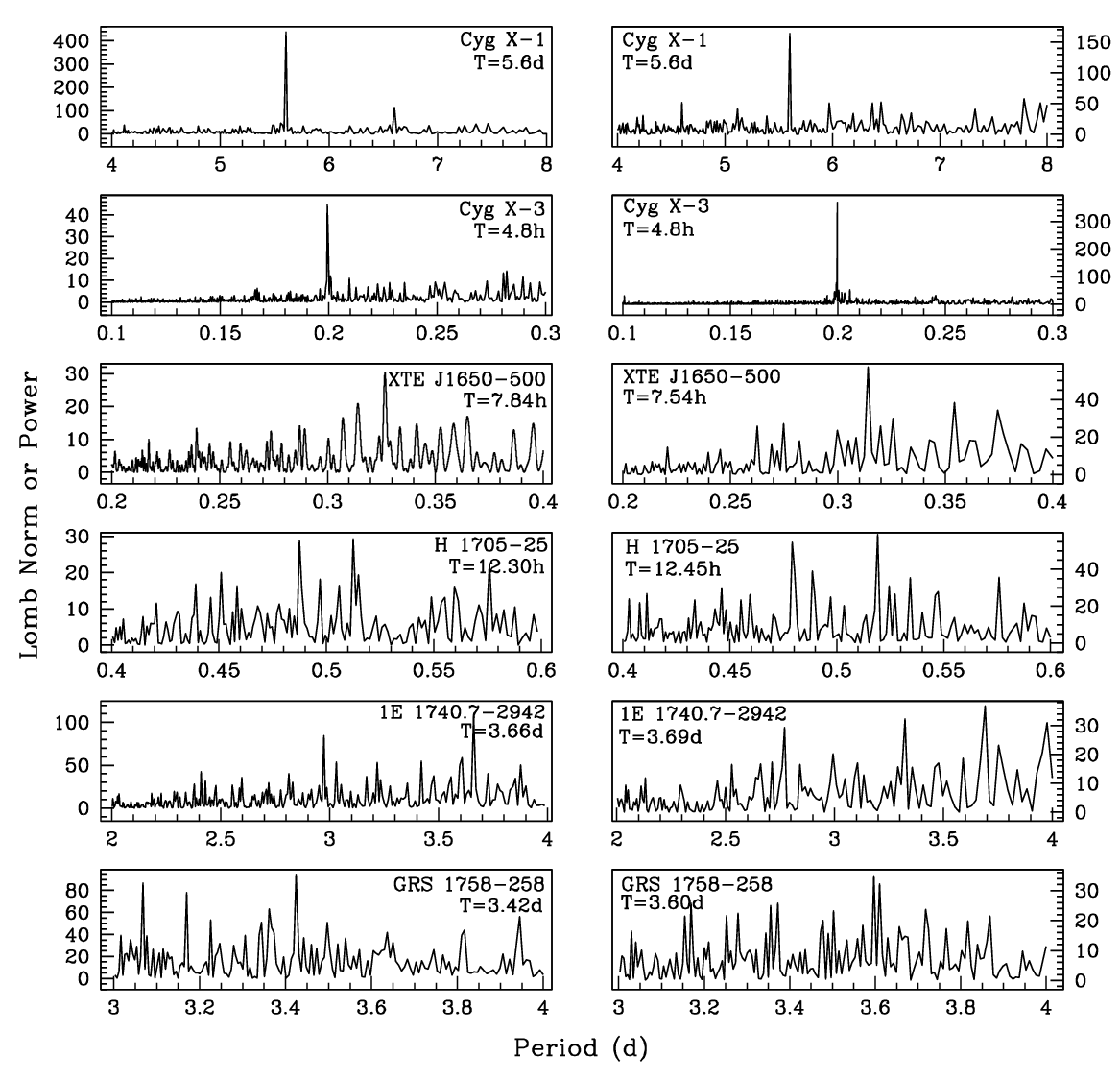}
\caption{Periodograms with RXTE/ASM (left) and Swift/BAT (right) data. Relevant periods are written in each box.}
\end{center}
\end{figure}


\newpage


\end{document}